\documentclass{iaus}
\usepackage{graphicx}

\title[SF in bulges from GALEX] 
{Star Formation in Bulges from GALEX}

\author[Sukyoung K. Yi]   
{Sukyoung K. Yi$^1$}

\affiliation{$^1$Yonsei University, Department of Astronomy, Seoul 120-749, Korea \break email:yi@yonsei.ac.kr}

\pubyear{2007}
\volume{245}  
\pagerange{1--6}
\date{?? and in revised form ??}
\jname{Proceedings Title IAU Symposium}
\editors{A.C. Editor, B.D. Editor \& C.E. Editor, eds.}
\begin{document}

\maketitle

\begin{abstract}
Early-type galaxies, considered as large bulges, have been found to have had
a much-more-than-boring star formation history in recent years by the UV
satellite GALEX. The most massive bulges, brightest cluster galaxies,
appear to be relatively free of young stars. But smaller bulges, normal ellipticals and lenticulars, often show unambiguous sign of recent star formation in their UV flux. The fraction of such UV-bright bulges in the volume-limited sample climbs up to the staggering 30\%. The bulges of spirals follow similar trends but a larger fraction showing signs of current and recent star formation. The implication on the bulge formation and evolution is discussed.
\end{abstract}

\firstsection 
\section{Introduction}

Bulges, mainly early-type galaxies and spiral bulges, are usually believed to
be predominantly composed of old stars and hence were not initially considered
as a primary science target for the UV survey mission, GALEX. The ``dying''
populations were simply not expected to exhibit a dramatic feature in the UV.
But GALEX broke that prejudice. Many bulges are found to radiate in the UV
much more than what can be explained by the theories of old stars, and that
includes even super-$L_*$, bright elliptical galaxies. This is intriguing
in many senses but most notably from the perspective of early-type galaxy
formation. It would be a cliche to mention yet another time
about the confrontation between the general belief based on simple observations (such as colour-magnitude relations and the fundamental plane) and the theoretical prediction from the hierarchical $\Lambda$CDM universe. An interesting implication can also be made on the role of supermassive black holes on the star formation history of bulges. Numerous groups are working on this issue with varieties of approaches, views or interpretations. I must review all to be fair but am forced to restrict to the works that I am involved in, considering that the field is still new and interpretation of the data is in the embryonic stage.


The GALEX satellite is equipped with two UV detectors: FUV and NUV centred on 1600\AA\ and 2500\AA, respectively (\cite[Martin et al. 2005]{martin05}). One of its strength is the wide field of view of 1.2\,degree. Its imaging operation works mainly in three modes in terms of exposure: the shallow ``all imaging survey'' (AIS), the medium-depth 1-orbit ``medium imaging survey'' (MIS), and the 20-orbit ``deep imaging survey'' (DIS). Our work is based on the MIS and DIS exclusively, whose limiting magnitudes (both in FUV and NUV) are 23(AB) and 26(AB), respectively. More detailed information on the GALEX mission can be found at \cite{martin05,morrissey05}.

\section{Residual star formation}

\begin{figure}
\begin{center}
\includegraphics[width=12cm]{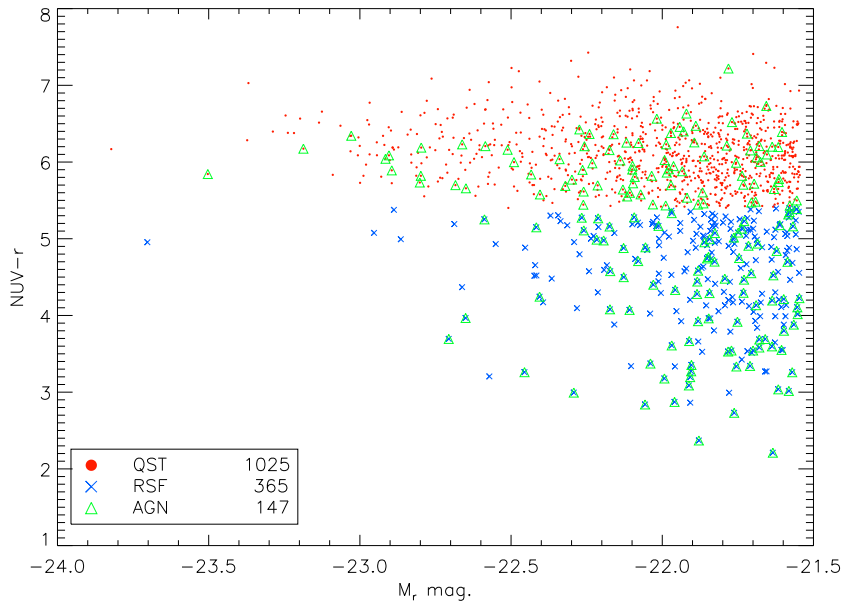}
  \caption{
  The NUV-optical colour-magnitude diagram. The horizontal dotted line at $NUV-r = 5.4$ is our residual-star-formation (RSF) vs quiescent criterion. We mark the ``quiescent'' galaxies with $NUV-r \geq 5.4$ with dots and the RSF galaxies with $NUV-r<5.4$ with crosses. Obscured AGN candidates and transition objects are shown as triangles.}\label{yi_f1}
\end{center}\end{figure}

The past efforts of understanding the residual star formation history in early-type galaxies were mostly based on optical data (e.g., \cite[Bower et al. 1992]{bower92}). The more powerful UV has been used (e.g., ANS, OAO-2, IUE, UIT, HUT, FOCA; see \cite[O'Connell 1999]{oconnell99} for review) to result in some insightful studies (e.g., \cite[Burstein et al. 1988]{burstein88}). The bottom line was that {\em early types are devoid of new star formation}. However, these surveys were done only to a small number of nearby and possibly heavily-biased sample of early-type galaxies. The seemingly-immortal IUE  observed only 31 early-type galaxies, while the others each observed fewer than 10. Such samples were hardly statistically significant while highly susceptible to selection biases, making the conclusions vulnerable to doubts.

With GALEX and SDSS we finally have a substantially large database spanning a large range in varieties of properties. The sample is hardly biased in any direction, either. While various sampling criteria are possible, the MIS mode allows us to reach passively-evolving red elliptical galaxies of roughly $L_*$ brightness ($M_r = -21.5$) out to $z=0.1$. In order to be free from the luminosity (Malmquist) bias we construct a volume-limited sample of roughly 2000 early-type ($fracDev \geq 0.95$) galaxies of $M_r \leq -21.5$ at $0.05 \leq z \leq 0.1$. A visual inspection removed about 30\% of the sample: some  appear to be late-type contaminants while others had a double detection within a short projected distance possibly posing a detection confidence problem. Readers are referred to \cite{kaviraj07}, \cite{schawinski07}, \cite{rich05} for more details. Since some significant fraction of early types are known to exhibit AGN activities that might contribute to the UV flux, we identify obscured AGN candidates by the Baldwin-Philips-Terlevich (BPT) test (1981).

Fig. 1 shows the NUV-optical colour-magnitude diagram of our sample of $\sim 1300$ galaxies. Their optical CMR is as tight as has always been known (see \cite[Kaviraj et al. 2007]{kaviraj07} and \cite[Schawinski et al. 2007]{schawinski07}), but the NUV CMR shows an order of magnitude larger scatter. The horizontal dotted line at $NUV-r = 5.4$ marks the colour of the nearby quiescent giant elliptical galaxy NGC\,1399. We assume that purely old populations may exhibit a UV flux as high as this. Based on the BPT test with galaxies with $S/N>3$ at least for [OIII] and [NII], roughly 30\% of our sample are classified as obscured AGN candidates. This figure changes to 15\% when we require all four emission lines to satisfy the S/N cut. As a result, $\sim 30$\% of our luminous early types had 1-3\% of their stellar mass formed in the last billion years or so. A naive estimation would suggest that all ellipticals formed $\sim 10$\% of their stars after $z=1$. {\em Ellipticals may have been forming stars all these years.}

\section{AGN feedback}

\begin{figure}
\begin{center}
\includegraphics[width=10cm,angle=0]{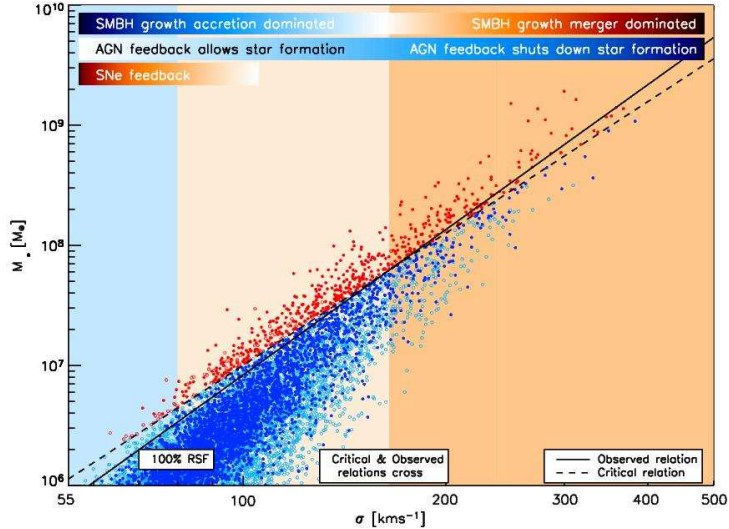}
\caption{
A schematic view of how the critical supermassive black hole mass - velocity dispersion relation regulates galaxy evolution. The full points are early-type galaxies, while empty circles are late-type galaxies. We show the observed $M_{\bullet}-\sigma$ relation (solid line) together with the best {\em critical} relation (dashed). We indicate the various regimes by different shades. The symbols above and below the dotted line are quiescent and RSF galaxies, respectively. Adopted from Schawinski et al. (2006).}\label{yi_f2}
\end{center}\end{figure}

It is not surprising at all that bright bulges have been forming stars. They are after all the most massive galaxies that are most likely to attract gases from outside and to internally retain the gas from the stellar mass loss. A more appropriate question would be {\em why are they not forming stars more rigorously?} Focusing on the empirical finding that all bulges have a supermassive black hole at its centre, much attention has been paid on the role of AGN feedback since the herald of \cite{dekel86}. These black holes may interact with their host galaxies by means of feedback in the form of energy and material jets; this feedback affects the evolution of the host and gives rise to observed relations between the black hole and the host \cite[Tremaine et al. (2002)]{tremaine02}.

In the GALEX data \cite{schawinski07} found a tendency that the UV-strong RSF galaxy fraction is much higher in the low-$\sigma$ (velocity dispersion) galaxies: 10\% for the most massive galaxies ($<\sigma>\sim 280\ \rm km/s$), and 30\% for lighter galaxies ($<\sigma>\sim 140$). Motivated by the tight correlation between the black hole mass $M_{BH}$ and $\sigma$, we hypothesized and derived an empirical relation for a {\em critical} black-hole mass for a given velocity dispersion above which the outflows from these black holes suppress star formation in their hosts by heating and expelling all available cold gas. Fig. 2 shows the schematic illustration of this hypothesis. The galaxies and their black hole masses are depicted following a single power law {\em a la Tremaine et al.} with a power 4 and the empirical scatter. Our {\em critical} relation for the feedback quenching is shown as a line with a slightly shallower slope. The galaxies above this line would have a black hole that is large enough to quench any residual star formation in the statistical sense. In this hypothesis, supermassive black holes are negligible in mass compared to their hosts but nevertheless play a critical role in the star formation history of galaxies.

\section{Bright cluster ellipticals: UV upturn}

\begin{figure}
\begin{center}
\includegraphics[width=10cm]{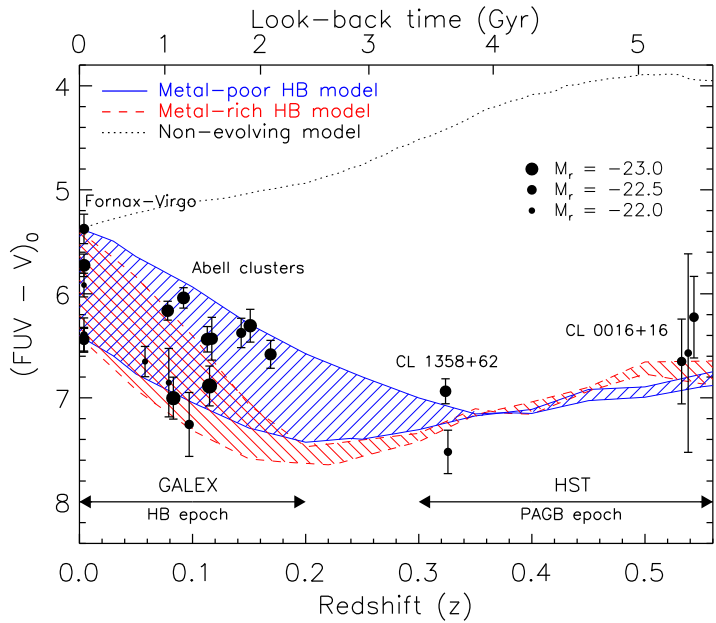}
  \caption{The UV upturn strength traced back in time using GALEX. The models are from Yi et al. (1999). The GALEX data symbols are scaled by the galaxy luminosity. From Ree et al. (2007).}\label{yi_f3}
\end{center}\end{figure}

Bright cluster galaxies (BCGs) are usually early-type cD galaxies. The local sample of BCGs (e.g., M87 and NGC\,1399) do not show much star formation activity and hence thought to be passive. By and large this is confirmed from our GALEX data analysis. These {\em quiescent} BCGs still show some UV flux mostly in the FUV, the phenomenon known as the UV upturn (e.g., \cite[O'Connell 1999]{oconnell99}). Their UV sources have been suspected to be evolved low-mass hot helium-burning stars. \cite{yi99} points out that their UV brightness would indicate much different ages for the stellar populations depending on {\em whether the UV sources are metal-rich or poor}. They suggested that this question can be answered by tracing the UV flux evolution in the past few billion years.

\cite{ree07} and \cite{lee05} have found 12 BCGs from GALEX MIS and DIS data at $0.05 < z < 0.17$ and the results are shown in Fig. 3. The FUV-optical colours are derived from the total integrated magnitudes. The nearby galaxy sample are shown at $z\approx 0$. The MIS and DIS data are shown as dots. The symbol size reflects the galaxy $V$-band luminosity. The two distant cluster data at $z>0.3$ are the HST/STIS data. The range of metal-rich and poor models shown in backward- and forward-hashed regions. The non-evolving model based on NGC\,1399 is shown in dotted line, and it is obviously not compatible with the observation. The old helium-burning star hypothesis is in excellent agreement, but it is still difficult to tell whether metal-rich or poor model reproduces the data better. It is nevertheless very promising and as we gather more data from GALEX the answer should be revealed.

\section{Lenticulars and Spiral bulges}

\begin{figure}
\begin{center}
 \includegraphics[width=12cm]{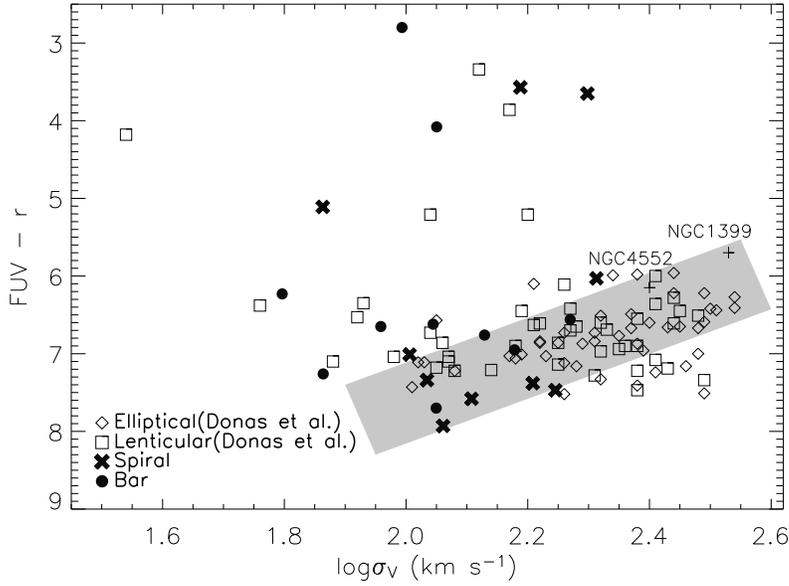}
  \caption{Spiral bulges compared to early-type galaxies. Early-type galaxy data are from Donas et al. (2007) and the spiral bulges are measured from 22 late-type galaxy images obtained by GALEX. The grey region roughly depicts the quiescent early-type galaxy sequence of Burstein et al. (1988). Note that a large fraction of spiral bulges are UV strong indicative of current/recent star formation. Errors are typically up to 0.3mag.}\label{yi_f4}
\end{center}\end{figure}

The early study of \cite{gregg89} found that some lenticular galaxies had a young stellar components mainly in their discs. \cite{proctor02} also suggest that lenticulars hint at distinct stellar populations from those of ellipticals from thir spectroscopic analysis. The presence of young population in lenticulars has also been suggested by \cite{donas07} and \cite{boselli05} using GALEX Nearby Galaxy Survey data (which operate in MIS mode). These studies unanimously announce that lenticulars have a larger variety in stellar age. The same data from \cite{donas07} are shown in Fig. 4. Ellipticals (by RC3) form a tight relation between the UV strength and velocity dispersion (similarly with Mg2 index). The majority of lenticulars also follow the relation but some 30\% do not, which can be most easily explained if they contain young stars. Also shown here are the bulges of spiral galaxies. We separate ordinary spirals and barred spirals here. Roughly half of the 22 late-type galaxies seriously depart from the early-type galaxy sequence, strongly suggesting that the stellar contents of spiral bulges are not exactly the same as those of early-type galaxies.

There is a debate on the origin of bulges. \cite{kormendy04} argued that classical bulges and peudobulges have markedly different origin and kinematic \& stellar properties. The GALEX analysis seems compatible for now while a more detailed analysis is in due course.

\section{Conclusion}

Prior to GALEX, a present star formation study on bulges was a taboo. Studies had to be based on a small number of sample galaxies mostly from the local universe. Now with GALEX, and other large surveys such as the SDSS, we finally have a statistically significant sample. The outcome is shocking. {\em Many mischievous bulges, even quite a few massive ones, have been secretly forming stars.} The residual star formation favours lower-mass bulges and perhaps lower-density environments as well. The residual star formation seems regulated perhaps by AGN feedback and the most massive bulges seem by and large quiescent. This still poses a challenge to theorists. Overly-simplistic scenario on the bulge formation, that asserts their universality, is no longer compelling.

\begin{acknowledgments}
This review is based on many people's hard work. Special thanks go to the GALEX science operation and data analysis team, Karl Forster, Mark Seibert, Todd Small, and Ted Wyder. Much of the work on the residual star formation has been performed by Sugata Kaviraj, Sadegh Khochfar, Kevin Schawinski, Hyunjin Jeong, Yun-Kyeong Sheen, Dowon Yi, \& Seok-Joo Joo. Works on the passive ellipticals have been led by Young-Wook Lee and Chang H. Ree. Lots of advice and criticism were presented by Alessandro Boselli, Jean-Michel Deharveng, Jose Donas, Mike Rich, and Samir Salim. GALEX is a NASA's small explorer mission funded by NASA. This work was supported by grant No. R01-2006-000-10716-0 from the Basic Research Program of the KOSEF.
\end{acknowledgments}

\newpage

\begin{discussion}

\discuss{Kormendy}{About the large scatter in the UV properties of spiral bulges, do you find any systematic difference between classical bulges and pseudobulges?
}

\discuss{Yi}{While a full investigation is under way, we find that peudobulges show more varieties in the UV properties, hinting for a wide spectrum in stellar age.
}

\discuss{Davies}{Regarding your comment that RSF galaxies account for half of the tilt of the fundamental plane slope and scatter, note that the SAURON studies find nearly all of the tilt and scatter could be explained by the presence of young stars in bulges.
}

\discuss{Yi}{
That could very well be the case. Our analysis is still preliminary and our estimate of 50\% is likely a lower limit.}

\discuss{Rich}{You claim that the Burstein et al. relation ($\sigma$ vs UV) is confirmed in your sample. But Rich et al. (2005) suggest otherwise. Where is the difference from?}

\discuss{Yi}{
I simply used Donas et al. measurements. It may have originated from sample selection but requires a further inspection.
}

\discuss{van der Wel}{
If you remove the RSF components from the RSF galaxies, would the remnant be consitent with the fundamental plane from the Virial theorem?
}

\discuss{Yi}{
A very good question. Unfortunately, the GALEX images have poor spatial resolution and the RSF component subtraction is not feasible for the galaxies in our sample. But that should be doing for nearby galaxies and would be an interesting investigation.
}

\end{discussion}
\end{document}